\documentclass[epjCONF,columns]{svjour} % for 2 columns format
\usepackage{graphics}
\usepackage[varg]{txfonts} % Times fonts
\usepackage[latin1]{inputenc}
\usepackage{bm}
\usepackage{xcolor}
\session-title{NSRT12}
\begin{document}
\title{Potential energy surfaces of actinide and transfermium nuclei from
       multi-dimensional constraint covariant density functional theories}
\author{Bing-Nan Lu\inst{1,2} \and
        Jie Zhao\inst{1} \and
        En-Guang Zhao\inst{1,3} \and
        Shan-Gui Zhou\inst{1,3}\fnmsep\thanks{\email{sgzhou@itp.ac.cn}}}
\institute{%
State Key Laboratory of Theoretical Physics, Institute of Theoretical
Physics, Chinese Academy of Sciences, Beijing 100190, China \and
Physics Department, Faculty of Science, University of Zagreb, 10000 Zagreb, Croatia \and
Center of Theoretical Nuclear Physics, National Laboratory of Heavy
Ion Accelerator, Lanzhou 730000, China
}
\abstract{
Multi-dimensional constrained covariant density functional theories were developed recently.
In these theories, all shape degrees of freedom $\beta_{\lambda\mu}$ deformations
with even $\mu$ are allowed, e.g., $\beta_{20}$, $\beta_{22}$, $\beta_{30}$, $\beta_{32}$,
$\beta_{40}$, $\beta_{42}$, $\beta_{44}$, and so on
and the CDFT functional can be one of the following four forms:
the meson exchange or point-coupling nucleon interactions combined with
the non-linear or density-dependent couplings.
In this contribution, some applications of these theories are presented.
The potential energy surfaces of actinide nuclei in the $(\beta_{20}, \beta_{22}, \beta_{30})$
deformation space are investigated.
It is found that besides the octupole deformation, the triaxiality also
plays an important role upon the second fission barriers.
The non-axial reflection-asymmetric $\beta_{32}$ shape in some transfermium
nuclei with $N=150$, namely $^{246}$Cm, $^{248}$Cf, $^{250}$Fm, and $^{252}$No
are studied.
} %end of abstract
\maketitle
\section{Introduction}
\label{sec:intro}

\begin{figure}
\begin{center}
\resizebox{0.99\columnwidth}{!}{%
 \includegraphics{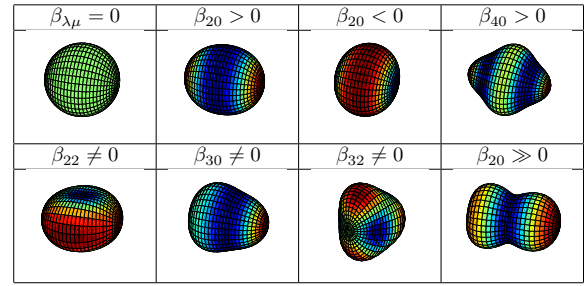} }
\end{center}
\caption{\label{Pic:deformations}(Color online)
A schematic show of some typical nuclear shapes: (a) sphere; (b) prolate spheroid;
(c) oblate spheroid; (d) hexadecapole shape; (e) triaxial ellipsoid;
(f) reflection symmetric octupole shape; (g) tetrahedron;
(h) reflection asymmetric octupole shape with very large quadrupole deformation.
Taken from Ref.~\cite{Lu2012_PhD}.
}
\end{figure}

``Shape'' gives an intuitive understanding of spatial
density distributions of quantum many-body systems including atomic nuclei.
For the description of the nuclear shape, it is convenient to
use the following parametrization
\begin{equation}
 \beta_{\lambda\mu} = {4\pi \over 3AR^\lambda} \langle Q_{\lambda\mu} \rangle,
 \label{eq:01}
\end{equation}
where $Q_{\lambda\mu}$ are the mass multipole operators.
A schematic show of some typical shapes is given in Fig.~\ref{Pic:deformations}~\cite{Lu2012_PhD}.
The majority of observed nuclear shapes is of spheroidal form which
is usually described by $\beta_{20}$.
Higher-order deformations with $\lambda>2$ such as $\beta_{30}$
also appear in certain atomic mass regions~\cite{Butler1996_RMP68-349}.
In addition, non-axial shapes in atomic nuclei, in particular,
the nonaxial-quadrupole (triaxial) deformation $\beta_{22}$ have been
studied both experimentally and
theoretically~\cite{Starosta2001_PRL86-971,Odegard2001_PRL86-5866,Meng2010_JPG37-064025}.
There is no a priori reason to neglect the nonaxial-octupole deformations,
especially the $\beta_{32}$
deformation~\cite{Hamamoto1991_ZPD21-163,Skalski1991_PRC43-140,Li1994_PRC49-R1250}.

Furthermore, more shape degrees of freedom play important roles
in the study of potential energy surfaces of atomic nuclei.
Particularly, various shape degrees of freedom play important and different roles
in the occurrence and in determining the heights of the inner and outer barriers
in actinide nuclei (in these nuclei double-humped fission barriers usually appear).
For example, the inner fission barrier is usually lowered when the triaxial
deformation is allowed, while for the outer barrier the reflection asymmetric
(RA) shape is favored~\cite{Pashkevich1969_NPA133-400,Moeller1970_PLB31-283,Girod1983_PRC27-2317,Rutz1995_NPA590-680,Abusara2010_PRC82-044303}.
Nowadays, it becomes more and more desirable to have accurate predictions of fission
barriers also for superheavy nuclei~\cite{Sobiczewski2007_PPNP58-292,Moeller2009_PRC79-064304,Pei2009_PRL102-192501,Xia2011_SciChinaPAM54S1-109,Liu2011_PRC84-031602R,Wang2012_PRC85-041601R}.
It is usually customary to consider only the triaxial and reflection symmetric (RS)
shapes for the inner barrier and axially symmetric and RA shapes for the outer
one~\cite{Moeller2009_PRC79-064304,Egido2000_PRL85-1198,Bonneau2004_EPJA21-391,Warda2012_PRC86-014322}.
The non-axial octupole deformations are considered in both the macroscopic-microscopic (MM)
models~\cite{Jachimowicz2011_PRC83-054302} and the non-relativistic
Hartree-Fock theories~\cite{Skalski2007_PRC76-044603}.

In order to give a microscopic and self-consistent study of
the potential energy surface with more shape degrees of freedom included,
multi-dimensional constrained covariant density functional theories are developed
recently~\cite{Lu2012_PRC85-011301R,Lu2012_in-prep}.
In these theories, all shape degrees of freedom $\beta_{\lambda\mu}$ deformations
with even $\mu$ are allowed, e.g., $\beta_{20}$, $\beta_{22}$, $\beta_{30}$, $\beta_{32}$,
$\beta_{40}$, $\beta_{42}$, $\beta_{44}$, and so on.
In this contribution, we present two recent applications of these theories:
the potential energy surfaces of actinide nuclei and
the non-axial reflection-asymmetric $\beta_{32}$ shape in some transfermium
nuclei.
In Section~\ref{sec:formalism}, the formalism of our multi-dimensional constrained
covariant density functional theories will be given briefly.
The results and discussions are presented in Section~\ref{sec:results}.
Finally we give a summary in Section~\ref{sec:summary}.

\section{Formalism}
\label{sec:formalism}

The details of the formalism for covariant density functional theories
can be found in
Refs.~\cite{Serot1986_ANP16-1,Reinhard1989_RPP52-439,Ring1996_PPNP37-193,Vretenar2005_PR409-101,Meng2006_PPNP57-470,Niksic2011_PPNP66-519}.
The CDFT functional in our multi-dimensional constrained calculations
can be one of the following four forms:
the meson exchange or point-coupling nucleon interactions combined with
the non-linear or density-dependent couplings~\cite{Lu2012_PRC85-011301R,Lu2012_in-prep}.
Here we show briefly the one corresponding to the non-linear point coupling
(NL-PC) interactions.
The Starting point of the relativistic
NL-PC density functional is the following Lagrangian:
\begin{equation}
 \mathcal{L} = \bar{\psi} \left( i\gamma_{\mu}\partial^{\mu}-M_{B} \right) \psi
             - \mathcal{L}_{{\rm lin}} -\mathcal{L}_{{\rm nl}}
             - \mathcal{L}_{{\rm der}} -\mathcal{L}_{{\rm cou}},
\end{equation}
where
\begin{eqnarray}
 \mathcal{L}_{{\rm lin}} & = & \frac{1}{2} \alpha_{S}\rho_{S}^{2}
                              +\frac{1}{2}\alpha_{V} j_{V}^{2}
                              +\frac{1}{2}\alpha_{TS}\vec{\rho}_{TS}^{2}
                              +\frac{1}{2}\alpha_{TV}\vec{j}_{TV}^{2}
 ,
 \nonumber \\
 \mathcal{L}_{{\rm nl }} & = & \frac{1}{3}\beta_{S}  \rho_{S}^{3}
                              +\frac{1}{4}\gamma_{S} \rho_{S}^{4}
                              +\frac{1}{4}\gamma_{V} \left[j_{V}^{2}\right]^{2}
 ,
 \nonumber \\
 \mathcal{L}_{{\rm der}} & = & \frac{1}{2}\delta_{S} \left[\partial_{\nu}\rho_{S}\right]^{2}
                              +\frac{1}{2}\delta_{V} \left[\partial_{\nu}j_{V}^{\mu}\right]^{2}
                              +\frac{1}{2}\delta_{TS}\left[\partial_{\nu}\vec{\rho}_{TS}\right]^{2}
 \nonumber \\
                         &   &+\frac{1}{2}\delta_{TV}\left[\partial_{\nu}\vec{j}_{TV}\right]^{2}
 ,
 \nonumber \\
 \mathcal{L}_{{\rm cou}} & = & \frac{1}{4} F^{\mu\nu}F_{\mu\nu}
                              +e \frac{1-\tau_{3}}{2}A_{\mu}j_{V}^{\mu}
 ,
\end{eqnarray}
are the linear, non-linear, and derivative couplings and the Coulomb part, respectively.
$M_{{\rm B}}$ is the nucleon mass, $\alpha_{S}$, $\alpha_{V}$, $\alpha_{TS}$,
$\alpha_{TV}$, $\beta_{S}$, $\gamma_{S}$, $\gamma_{V}$, $\delta_{S}$, $\delta_{V}$,
$\delta_{TS}$, and $\delta_{TV}$ are coupling constants for different channels and
$e$ is the electric charge.
$\rho_{S}$, $\rho_{TS}$, $j_{V}$, and $j_{TV}$ are the iso-scalar density,
iso-vector density, iso-scalar current, and iso-vector current, respectively.
The densities and currents are defined as:
\begin{eqnarray}
 \rho_{S} & = & \bar{\psi}\psi ,
 \\
 \rho_{TS} & = & \bar{\psi}\vec{\tau}\psi ,
 \\
 j_{V}^{\mu} & = & \bar{\psi}\gamma^{\mu}\psi \label{eq:jv},
 \\
 j_{TV}^{\mu} & = & \bar{\psi}\vec{\tau}\gamma^{\mu}\psi \label{eq:jtv}.
\end{eqnarray}
Starting from the above Lagrangian, using the Slater determinants as
trial wave functions and neglecting the Fock term as well as the contributions
to the densities and currents from the negative energy levels, one
can derive the equations of motion for the nucleons.
Furthermore, for systems with time reversal symmetry, only the time-like components 
of the vector currents (\ref{eq:jv}) and (\ref{eq:jtv}) survive.
The resulted equation for the nucleons reads
\begin{equation}
 \hat{h}\psi_{i} = \left(\bm{\alpha}\cdot\vec{p}+\beta(M+S(\vec{r}))+V(\vec{r})\right)\psi_{i}
                 = \epsilon_{i}\psi_{i}
 ,
\end{equation}
where the potentials $V(\bm{r})$ and $S(\bm{r})$ are calculated as
\begin{eqnarray}
S & = & \alpha_{S}\rho_{S}+\beta_{S}\rho_{S}^{2}+\gamma_{S}\rho_{S}^{3}+\delta_{S}\triangle\rho_{S}\nonumber \\
 &  & +\left(\alpha_{TS}\rho_{TS}+\delta_{TS}\triangle\rho_{TS}\right)\tau_{3}  ,\\
V & = & \alpha_{V}\rho_{V}+\gamma_{V}\rho_{V}^{3}+\delta_{V}\triangle\rho_{V}W\nonumber \\
 &  & +\left(\alpha_{TV}\rho_{TV}+\delta_{TV}\triangle\rho_{TV}\right)\tau_{3}  .
\end{eqnarray}

An axially deformed harmonic oscillator (ADHO) basis is adopted for solving the
Dirac equation~\cite{Lu2012_PRC85-011301R,Lu2012_in-prep,Lu2011_PRC84-014328}.
The ADHO basis are defined as the eigen solutions of the Schrodinger
equation with an ADHO potential~\cite{Gambhir1990_APNY198-132,Ring1997_CPC105-77},
\begin{eqnarray}
 \left[ -\frac{\hbar^{2}}{2M} \nabla^{2} + V_{B}(z,\rho) \right] \Phi_{\alpha}(\bm{r}\sigma)
 & = &
 E_{\alpha} \Phi_{\alpha}(\bm{r}\sigma)
 ,
 \label{eq:BasSchrodinger-1}
\end{eqnarray}
where
\begin{equation}
 V_{B}(z,\rho) = \frac{1}{2} M ( \omega_{\rho}^{2}\rho^{2} + \omega_{z}^{2}z^{2} )
 ,
\end{equation}
is the axially deformed HO potential and $\omega_{z}$ and $\omega_{\rho}$
are the oscillator frequencies along and perpendicular to $z$ axis, respectively.
The solution of Eq.~(\ref{eq:BasSchrodinger-1}) reads
\begin{equation}
 \Phi_{\alpha}(\bm{r}\sigma) =
  C_{\alpha} \phi_{n_{z}}(z) R_{n_{\rho}}^{m_{l}}(\rho)
  \frac{1}{\sqrt{2\pi}} e^{im_{l}\varphi} \chi_{s_{z}}(\sigma),
\end{equation}
where $\phi_{n_{z}}(z)$ and $R_{n_{\rho}}^{m_{l}}(\rho)$ are the
HO wave functions,
\begin{eqnarray}
 \phi_{n_{z}}(z) & = &
  \frac{1}{\sqrt{b_{z}}} \frac{1}{\pi^{{1}/{4}} \sqrt{2^{n_{z}}n_{z}!}}
  H_{n_{z}} \left(\frac{z}{b_{z}}\right) e^{-{z^{2}}/{2b_{z}}}
 ,\nonumber \\
 R_{n_{\rho}}^{m_{l}}(\rho) & = &
  \frac{1}{b_{\rho}} \sqrt{\frac{2n_{\rho}!}{(n_{\rho}+|m_{l}|)!}}
  \left( \frac{\rho}{b_{\rho}} \right)^{|m_{l}|}
  L_{n_{\rho}}^{|m_{l}|}\left(\frac{\rho^{2}}{b_{\rho}^{2}}\right)
  e^{-{\rho^{2}}/{2b_{\rho}^{2}}}
 ,\nonumber \\
\end{eqnarray}
$\chi_{s_{z}}$ is a two component spinor and $C_{\alpha}$ is a complex
number inserted for convenience.
Oscillator lengths $b_{z}$ and $b_{\rho}$ are related to the frequencies
by $b_{z}=1/\sqrt{M\omega_{z}}$ and $b_{\rho}=1/\sqrt{M\omega_{\rho}}$.
%The cooresponding eigen energy is
%$E_{\alpha}=\omega_{\rho}(2n_{\rho}+|m_{l}|+1)+\omega_{z}(n_{z}+\frac{1}{2})$
%and the major quantum number is $N_{\alpha}=2n_{\rho}+|m_{l}|+n_{z}$.

These basis are also eigen functions of the $z$ component of the
angular momentum $j_{z}$ with eigen values $K=m_{l}+m_{s}$.
For any basis state $\Phi_{\alpha}(\bm{r}\sigma)$, the time reversal state
is defined as $\Phi_{\bar{\alpha}}(\bm{r}\sigma)=\mathcal{T}\Phi_{\alpha}(\bm{r}\sigma)$,
where $\mathcal{T}=i\sigma_{y}K$ is the time reversal operator and
$K$ is the complex conjugation.
Apparently we have $K_{\bar{\alpha}}=-K_{\alpha}$
and $\pi_{\bar{\alpha}}=\pi_{\alpha}$.
These basis form a complete set for expanding any two-component spinors.
For a Dirac spinor with four components,
\begin{equation}
 \psi_{i}(\bm{r}\sigma) =
 \left( \begin{array}{c}
        \sum_{\alpha}f_{i}^{\alpha} \Phi_{\alpha}(\bm{r}\sigma) \\
        \sum_{\alpha}g_{i}^{\alpha} \Phi_{\alpha}(\bm{r}\sigma)
        \end{array}
 \right),
\end{equation}
where the sum runs over all the possible combination of the quantum
numbers $\alpha=\{n_{z},n_{r},m_{l},m_{s}\}$, and $f_{i}^{\alpha}$ and
$g_{i}^{\alpha}$ are the expansion coefficients.
In practical calculations, one should truncate the basis
in an effective way~\cite{Lu2012_PRC85-011301R,Lu2012_in-prep,Lu2011_PRC84-014328}.

We expand the potentials $V(\bm{r})$ and $S(\bm{r})$ and various densities
in terms of the Fourier series,
\begin{equation}
 f(\rho,\varphi,z) =
 \sum_{\mu=-\infty}^{\infty} f_{\mu}(\rho,z) \frac{1}{\sqrt{2\pi}} \exp(i\mu\varphi) .
 \label{eq:Fourier}
\end{equation}
The nucleus is assumed to be symmetric under the $V_4$ group, that is, for all the 
potentials and densities we have
\begin{equation}
 f(\rho,\pi\pm\varphi,z)  = f(\rho,2\pi-\varphi,z) = f(\rho,\varphi,z).
\end{equation}
Thus the components $f_\mu$'s satisfy $f_{\mu}=f_{\mu}^{*}=f_{\bar{\mu}}$ and all the terms
with odd $\mu$ vanish.
The expansion Eq.~(\ref{eq:Fourier}) can be simplified as
\begin{equation}
 f(\rho,\varphi,z) =  f_{0}(\rho,z) \frac{1}{\sqrt{2\pi}}
+ \sum_{n=1}^{\infty} f_{n}(\rho,z) \frac{1}{\sqrt{\pi}}\cos(2n\varphi),
\end{equation}
where
\begin{eqnarray}
 f_{0}(\rho_{,}z) & = & \frac{1}{\sqrt{2\pi}}\int_{0}^{2\pi}d\varphi f(\rho,\varphi,z) ,
 \nonumber \\
 f_{n}(\rho,z)    & = & \frac{1}{\sqrt{ \pi}}\int_{0}^{2\pi}d\varphi f(\rho,\varphi,z)\cos(2n\varphi)
 ,
\end{eqnarray}
are real functions of $\rho$ and $z$.

The total energy of a nucleus reads
\begin{eqnarray}
 E_{{\rm total}} & = &
 \int d^{3}\bm{r}
  \left\{
   \sum_{k}v_{k}^{2} \psi_{k}^{\dagger} \left( \bm{\alpha}\cdot\bm{p}+\beta M \right) \psi_{k}\right.
 \nonumber \\
 &  & +\frac{1}{2}\alpha_{S}\rho_{S}^{2}
      +\frac{1}{2}\alpha_{V}\rho_{V}^{2}
      +\frac{1}{2}\alpha_{TS}\rho_{TS}^{2}
      +\frac{1}{2}\alpha_{TV}\rho_{TV}^{2}
 \nonumber \\
 &  & +\frac{1}{3}\beta_{S}\rho_{S}^{3}
      +\frac{1}{4}\gamma_{S}\rho_{S}^{4}
      +\frac{1}{4}\gamma_{V}\rho_{V}^{4}
 \nonumber \\
 &  & +\frac{1}{2}\delta_{S}\rho_{S}\Delta\rho_{S}
      +\frac{1}{2}\delta_{V}\rho_{V}\Delta\rho_{V}
 \nonumber \\
 &  & +\frac{1}{2}\delta_{TS}\rho_{TS}\Delta\rho_{TS}
      +\frac{1}{2}\delta_{TV}\rho_{TV}\Delta\rho_{TV}
 \nonumber \\
 &  & \left.+\frac{1}{2}e\rho_{C}A+E_{{\rm pair}}+E_{{\rm c.m.}}\right\} ,
\end{eqnarray}
where the center of mass correction $E_{{\rm c.m.}}$ can be calculated
either phenomenologically or microscopically.

The intrinsic multipole moments are calculated from the vector densities by
\begin{equation}
 Q^{\tau}_{\lambda\mu} = \int d^{3}\bm{r} \rho_{V} (\bm{r})r^{\lambda}Y_{\lambda\mu}(\Omega),
\end{equation}
where $Y_{\lambda\mu}(\Omega)$ is the spherical harmonics and $\tau$
refers to the proton, neutron or the whole nucleus.

The potential energy surface (PES) is obtained by the constrained self-consistent calculation,
\begin{equation}
 E^{\prime} = \langle \hat{H} \rangle
            + \sum_{n=1}^{N_{c}} \frac{1}{2} C_{n}
              \left( \langle \hat{Q}_{n} \rangle - \mu_{n}
              \right)^{2}
 ,
\end{equation}
where $\hat{H}$ is the RMF Hamiltonian, 
$\hat{Q}_{n}$'s are the multipole operators to be constrained and $N_{c}$ is the dimension of the constraining space.

Both the BCS approach and the Bogoliubov transformation are implemented
in our model to take into account the pairing effects.
For the pairing force, we can use a delta force or a separable finite-range
pairing force~\cite{Tian2009_PLB676-44,Tian2009_PRC80-024313,Niksic2010_PRC81-054318}.
More details of the multi-dimensional constraint covariant density
functional theories can be found in Refs.~\cite{Lu2012_PRC85-011301R,Lu2012_in-prep}.

\section{Results and discussions}
\label{sec:results}

\subsection{One-, two-, and three-dimensional potential energy surface of $^{240}$Pu}
\label{sec:240Pu}

\begin{figure}
%\begin{centering}
%\includegraphics[width=0.9\columnwidth]{pic/Pu240PES1d}
%\includegraphics[width=0.9\columnwidth]{Pu240PES1d}
%\par\end{centering}
\begin{center}
\resizebox{0.95\columnwidth}{!}{%
 \includegraphics{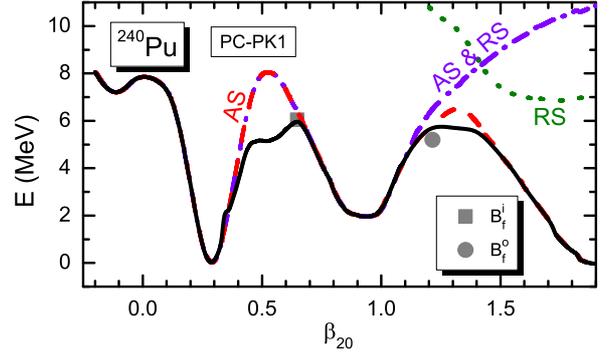} }
\end{center}
\caption{\label{Pic:PU240-1d}(Color online)
Potential energy curves of $^{240}$Pu with various self-consistent symmetries imposed.
The solid black curve represents the calculated fission path with $V_4$ symmetry imposed:
the red dashed curve that with axial symmetry (AS) imposed,
the green dotted curve that with reflection symmetry (RS) imposed,
the violet dot-dashed line that with both symmetries (AS \& RS) imposed.
The empirical inner (outer) barrier height $B_\mathrm{emp}$ is denoted by the grey square (circle).
The energy is normalized with respect to the binding energy of the ground state.
The parameter set used is PC-PK1.
Taken from Ref.~\cite{Lu2012_PRC85-011301R}.
}
\end{figure}

In Ref.~\cite{Lu2012_PRC85-011301R}, one- (1-d), two- (2-d), and three-dimensional (3-d)
constrained calculations were performed for the actinide nucleus $^{240}$Pu.
The parameter set PC-PK1 is used~\cite{Zhao2010_PRC82-054319}.
In Fig.~\ref{Pic:PU240-1d} we show the 1-d potential energy curves (PEC)
from an oblate shape with $\beta_{20}$ about $-0.2$ to the fission
configuration with $\beta_{20}$ beyond 2.0
which are obtained from calculations with different self-consistent symmetries
imposed: the axial (AS) or triaxial (TS) symmetries combined with
reflection symmetric (RS) or asymmetric cases.
The importance of the triaxial deformation on the inner barrier and
that of the octupole deformation on the outer barrier
are clearly seen:
The triaxial deformation reduces the inner barrier height by more than 2 MeV
and results in a better agreement with the empirical datum;
the RA shape is favored beyond the fission isomer and lowers very much
the outer fission barrier.
Besides these features, it was found for the first time that the outer
barrier is also considerably lowered by about 1 MeV when the triaxial
deformation is allowed.
In addition, a better reproduction of the empirical barrier height can be seen for
the outer barrier.
It has been stressed that this feature can only be found when the axial and
reflection symmetries are simultaneously broken~\cite{Lu2012_PRC85-011301R}.

\begin{figure}
%\begin{centering}
%\includegraphics[width=0.8\columnwidth]{pic/Pu240b2b3}
%\includegraphics[width=0.8\columnwidth]{Pu240b2b3}
%\par\end{centering}
\begin{center}
\resizebox{0.98\columnwidth}{!}{%
 \includegraphics{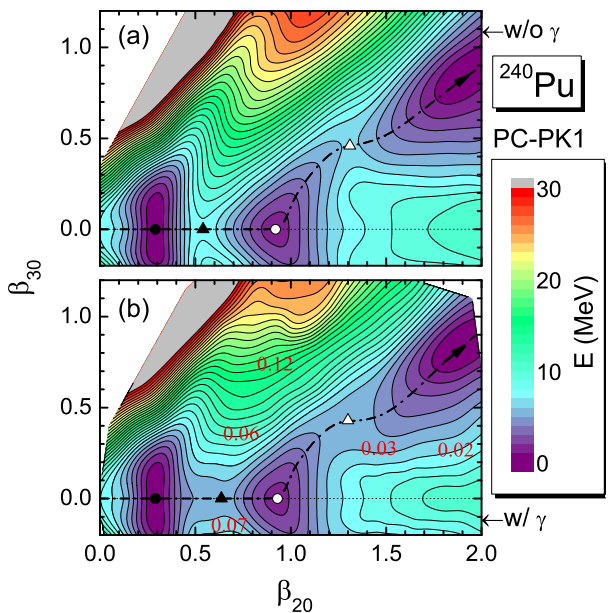} }
\end{center}
\caption{\label{Pic:PU240_2d}(Color online)
Potential energy surfaces of $^{240}$Pu in the $(\beta_{20},\beta_{30})$ plane
from calculations (a) without and (b) with the triaxial deformation included.
The energy is normalized with respect to the binding energy of the ground state.
The numbers in (b) show the values of $\beta_{22}$ at these points.
The fission path is represented by a dash-dotted line.
The ground state and fission isomer are denoted by full and open circles.
The first and second saddle points are denoted by full and open triangles.
The contour interval is 1 MeV.
Taken from Ref.~\cite{Lu2012_PRC85-011301R}.
}
\end{figure}

In order to see how the PES of $^{240}$Pu becomes unstable against
the triaxial distortion, 2-d PES's from calculations without and with
the triaxial deformation were compared in Fig.~\ref{Pic:PU240_2d}~\cite{Lu2012_PRC85-011301R}.
When the triaxial deformation is allowed, the binding energy of $^{240}$Pu
assumes its lowest possible value at each $(\beta_{20},\beta_{30})$ point.
At some points, especially those around the two saddle points,
non-axial solutions are favored than the axial ones.
The inner barrier height is lowered by about 2 MeV.
About 1 MeV is gained for the binding energy at the second saddle point
due to the triaxiality.
In the regions around the ground state and in the fission isomer valleys,
only axially symmetric solutions are obtained.

\begin{figure}
%\begin{centering}
%\includegraphics[width=0.8\columnwidth]{pic/PU240section2}
%\includegraphics[width=0.8\columnwidth]{PU240section2}
%\par\end{centering}
\resizebox{0.98\columnwidth}{!}{%
 \includegraphics{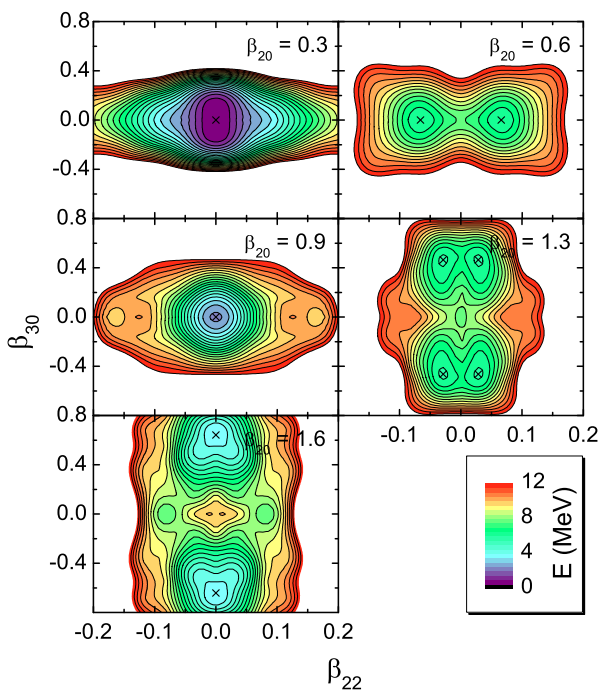} }
\caption{\label{Pic:PU240_3d}(Color online)
Sections of the three-dimensional PES of $^{240}$Pu in the
$(\beta_{22},\beta_{30})$ plane calculated at
$\beta_{20}$= 0.3 (around the ground state),
0.6 (around the first saddle point), 0.9 (around the fission isomer),
1.3 (around the second saddle point) and 1.6 (beyond the outer barrier), respectively.
The energy is normalized with respect to the binding energy of the ground state.
The contour interval is 0.5 MeV.
Local minima are denoted by crosses.
Taken from Ref.~\cite{Lu2012_PRC85-011301R}.
}
\end{figure}

A full 3-d PES has been obtained for $^{240}$Pu~\cite{Lu2012_PRC85-011301R}.
In Fig.~\ref{Pic:PU240_3d} are shown only five typical sections of 
the 3-d PES of $^{240}$Pu in the $(\beta_{22},\beta_{30})$ plane calculated
around the ground state, the first saddle point, the fission isomer,
the second saddle point, at a point beyond the outer barrier, respectively.
The following conclusions were drawn by examining these 3-d PES's~\cite{Lu2012_PRC85-011301R}:
(1) The ground state and the fission isomer are both axially and
reflection symmetric.
The stiffness of the fission isomer is much larger than that
of the ground state against both the $\beta_{22}$ and $\beta_{30}$ distortions.
(2) The second saddle point which is close to
$\beta_{20}=1.3$ appears as both triaxial and reflection asymmetric shape.
(3) The triaxial distortion appears only on the top of the fission barriers.

From the investigation of the one-, two-, and three-dimensional 
PES of $^{240}$Pu, we can learn a lot about the importance of different
shape degrees of freedom in different regions of PES in actinide nuclei.
These information could be useful in further systematic calculations.

\subsection{Inner and outer fission barriers of even-even actinide nuclei}
\label{sec:Bf}

Guided by the features found in the 1-d, 2-d, and 3-d PES's of $^{240}$Pu,
the fission barrier heights were extracted for even-even actinide
nuclei~\cite{Lu2012_PRC85-011301R}.
The calculated values were compared with empirical ones recommended
in RIPL-3~\cite{Capote2009_NDS110-3107}.

\begin{figure}
%\begin{centering}
%\includegraphics[width=0.7\columnwidth]{pic/actnidesFB}
%\includegraphics[width=0.7\columnwidth]{actnidesFB}
%\par\end{centering}
\begin{center}
\resizebox{0.95\columnwidth}{!}{%
 \includegraphics{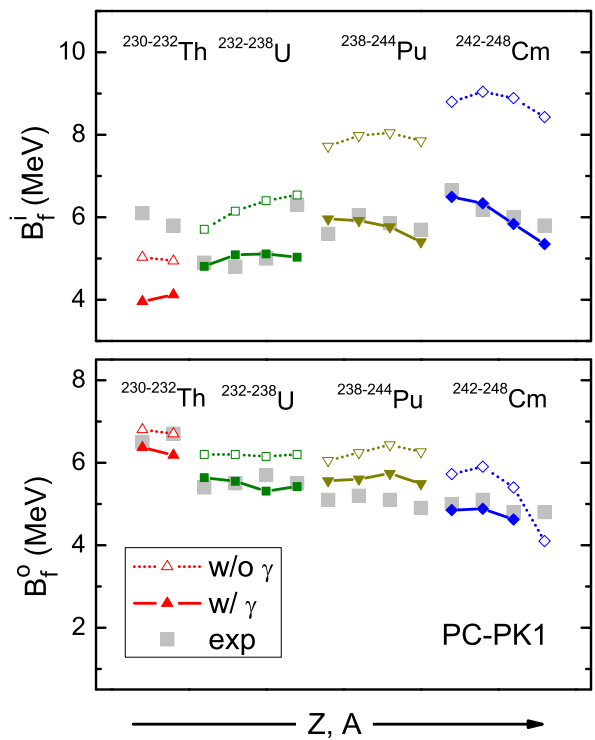} }
\end{center}
\caption{\label{fig:The-calculated-inner}(Color online)
The inner ($B^\mathrm{i}_\mathrm{f}$) and outer ($B^\mathrm{o}_\mathrm{f}$)
barrier heights of even-even actinide nuclei.
The axial (triaxial) results are denoted by open (full) symbols.
The empirical values are taken from Ref.~\protect\cite{Capote2009_NDS110-3107}
and represented by grey squares.
Taken from Ref.~\cite{Lu2012_PRC85-011301R}.
}
\end{figure}

As it has been shown previously, around the inner barrier an actinide nucleus
assumes triaxial and reflection symmetric shapes.
Thus in order to obtain the inner fission barrier height we can safely
make a one-dimensional constraint calculation with
the triaxial deformation allowed and the reflection symmetry imposed.
In Fig.~\ref{fig:The-calculated-inner}(a) we show the calculated inner barrier
heights $B^\mathrm{i}_\mathrm{f}$ and compare them with the empirical values.
It is seen that the triaxiality lowers the inner barrier heights
of these actinide nuclei by $1 \sim 4$ MeV
as what has been shown in Ref.~\cite{Abusara2010_PRC82-044303}.
In general the agreement of our calculation results with the empirical ones
is very good with exceptions in the two thorium isotopes and $^{238}$U.
Possible reasons for these disagreements were discussed in Ref.~\cite{Lu2012_PRC85-011301R}.

To obtain the outer fission barrier height $B^\mathrm{o}_\mathrm{f}$,
the situation becomes more complicated because more shape degrees of freedom
play important roles around the outer fission barrier.
In Ref.~\cite{Lu2012_PRC85-011301R}, 2-d constraint calculations
were made carefully around the second saddle points for even-even actinide nuclei.
In the lower panel of Fig.~\ref{fig:The-calculated-inner} we show the
results of outer barrier heights $B^\mathrm{o}_\mathrm{f}$ and compare
them with empirical values.
For most of the nuclei investigated here, the triaxiality lowers the outer barrier
by 0.5 $\sim$ 1 MeV, accounting for about 10 $\sim$ 20\% of the barrier height.
One finds that the calculation with the triaxiality agrees well with
the empirical values and the only exception is $^{248}$Cm.
From the calculation with the axial symmetry imposed, the outer barrier height
of $^{248}$Cm is already smaller than the empirical value.
The reason for this discrepancy may be related to that there are two
possible fission paths beyond the first barrier~\cite{Lu2012_PRC85-011301R}.

In Ref.~\cite{Lu2012_PRC85-011301R}, it was
also examined the parameter dependency of the influence of triaxiality
on the outer fission barrier and
the lowering effect of the triaxiality on the outer fission barrier
was also observed when parameter sets other than PC-PK1 are used.

\subsection{Non-axial octupole shapes in $N=150$ isotones}
\label{sec:tetrahedral}

Nowadays the study of nuclei with $Z\sim 100$ becomes more and more important
because it not only reveals the structure of these nuclei themselves
but also provides significant information for superheavy
nuclei~\cite{Herzberg2006_Nature442-896,Zhang2011_PRC83-011304R,Zhang2012_PRC85-014324,Zhang2012_arxiv1208.1156}.
One of the relevant and interesting topics is how to explain the low-lying
$2^-$ states in some $N = 150$ even-even nuclei.
In these nuclei, the bandhead energy $E(2^-)$ of the lowest $2^-$ bands
is very low~\cite{Robinson2008_PRC78-034308}.
It is well accepted that the octupole correlation is responsible for it.
For example, a quasiparticle phonon model with octupole correlations included was
used to explain the excitation energy of the $2^-$ state of the
isotones with $N = 150$~\cite{Jolos2011_JPG38-115103}.
In Ref.~\cite{Chen2008_PRC77-061305R}, Chen et al. proposed that the non-axial
octupole $Y_{32}$-correlation results in the experimentally observed low-energy $2^-$
bands in the $N = 150$ isotones and the reflection asymmetric shell model calculations
reproduces well the experimental observables of these $2^-$ bands.
%It was also predicted that the strong nonaxial-octupole effect may persist
%up to the element 108~\cite{Chen2008_PRC77-061305R} and
%play a crucial role in determining the shell stability in
%even heavier nuclei~\cite{Chen2012_in-prep}.

\begin{figure}
\begin{center}
\resizebox{0.80\columnwidth}{!}{%
 \includegraphics{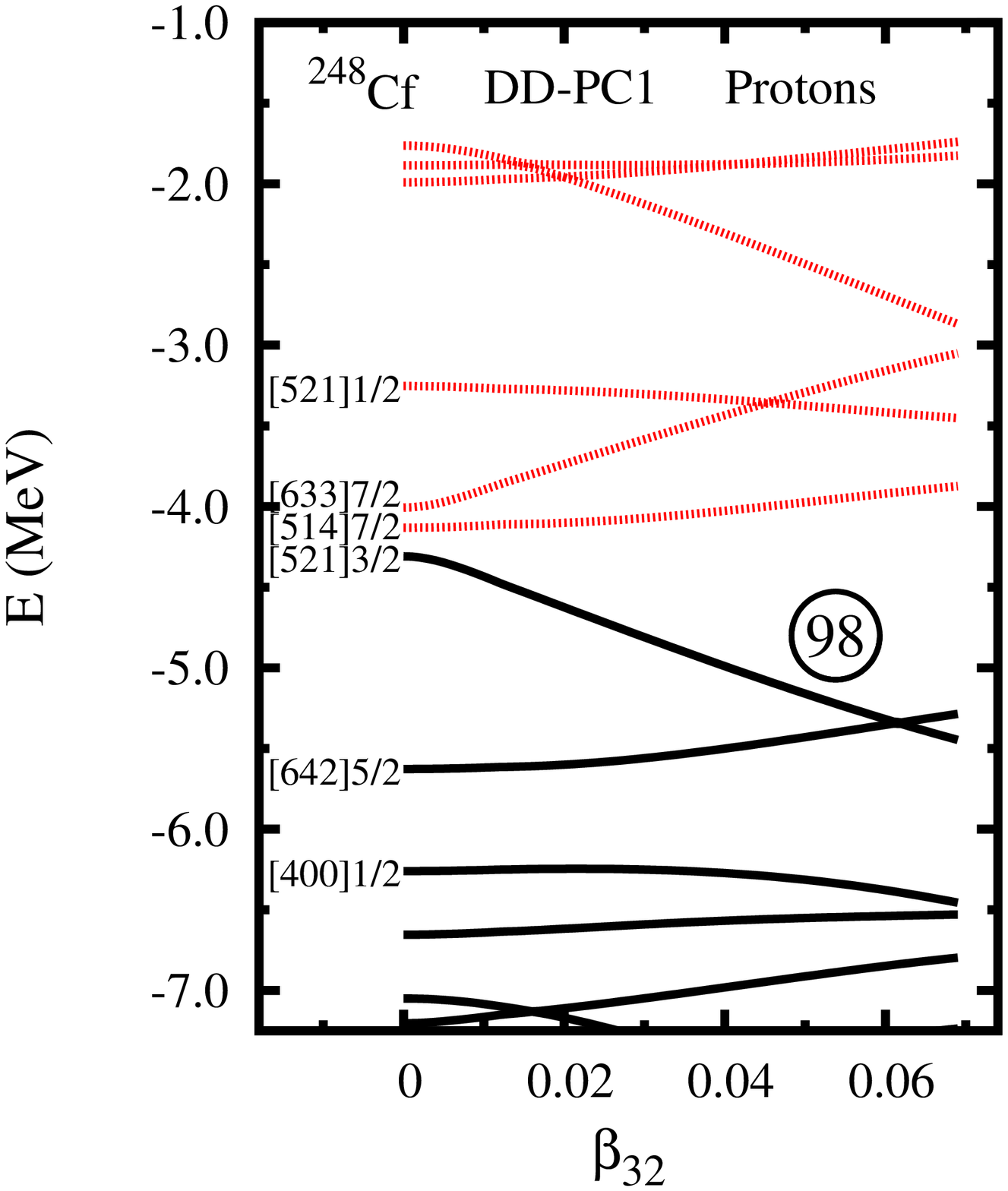} }
\resizebox{0.80\columnwidth}{!}{%
 \includegraphics{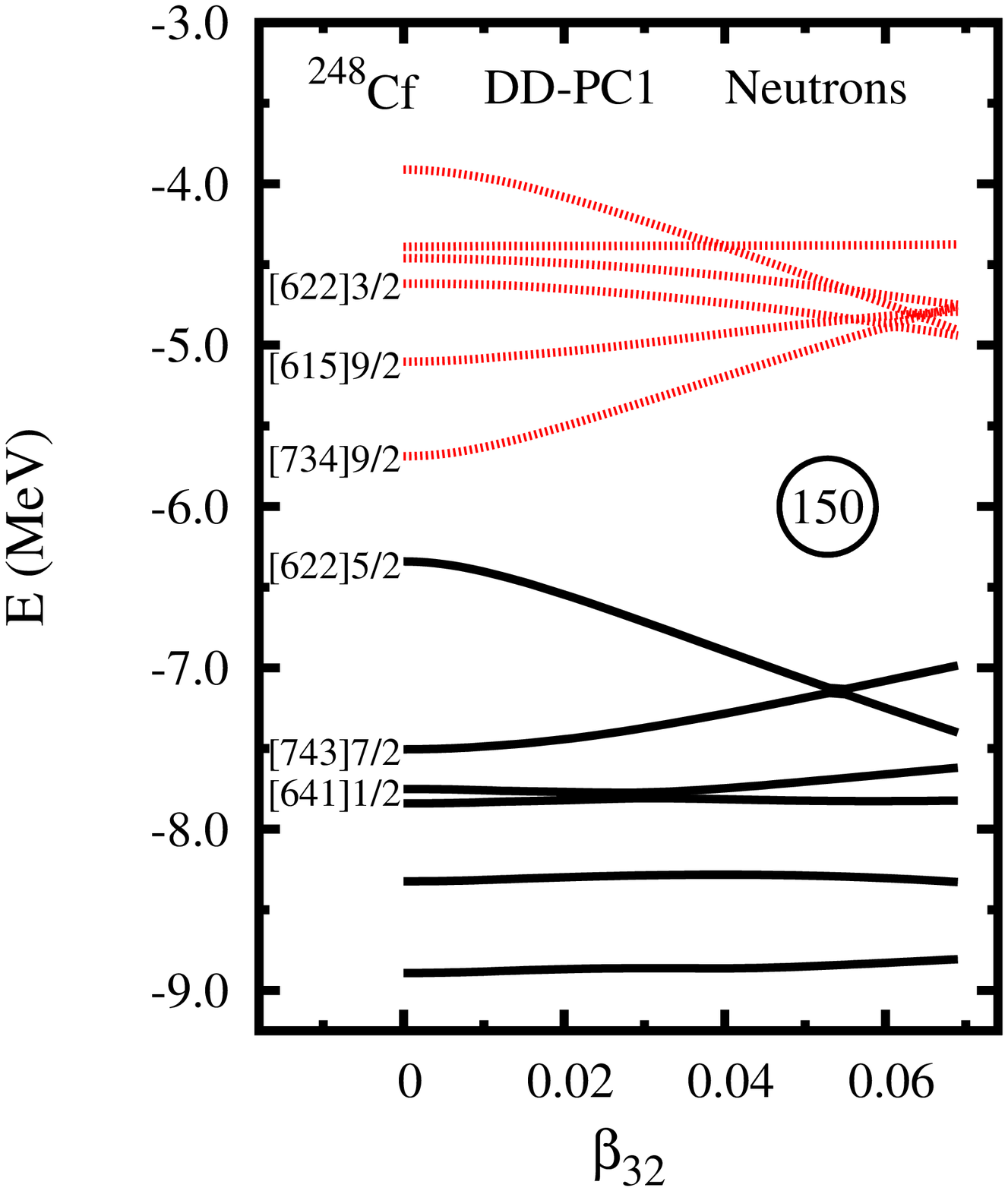} }
\end{center}
\caption{\label{fig:lev}(Color online)
The single-particle levels near the Fermi surface for (a) protons
and (b) neutrons of $^{248}$Cf as a function
of $\beta_{32}$ with $\beta_{20}$ fixed at 0.3.
}
\end{figure}

The non-axial reflection-asymmetric $\beta_{32}$ shape in some transfermium
nuclei with $N=150$, namely $^{246}$Cm, $^{248}$Cf, $^{250}$Fm, and $^{252}$No
were investigated with the multidimensional constrained covariant density functional
theory~\cite{Zhao2012_arxiv1209.6567}.
The parameter set DD-PC1 is used~\cite{Niksic2008_PRC78-034318}.
For the ground states of $^{248}$Cf and  $^{250}$Fm, the non-axial octupole deformation
parameter $\beta_{32} > 0.03$ and the energy gain due to the $\beta_{32}$
distortion is larger than 300 keV.
In $^{246}$Cm and $^{252}$No, shallow $\beta_{32}$ minima are found.

The triaxial octupole $Y_{32}$ effects stem from the coupling between pairs of
single-particle orbits with $\Delta j = \Delta l = 3$ and $\Delta K=2$
where $j$ and $l$ are the total and orbit angular momenta of single particles
respectively and $K$ the projection of $j$ on the $z$ axis.
In Fig.~\ref{fig:lev}, we show the proton and neutron single-particle levels
near the Fermi surface for $^{248}$Cf as a function
of $\beta_{32}$ with $\beta_{20}$ fixed at 0.3.
It was shown that the spherical proton orbitals $\pi 2f_{7/2}$ and $\pi 1i_{13/2}$ are very
close to each other~\cite{Zhao2012_arxiv1209.6567}.
This near degeneracy results in octupole correlations.
As seen in Fig.~\ref{fig:lev}, the two proton levels, $[521]3/2$
originating from $2f_{7/2}$ and $[633]7/2$
originating from $1i_{13/2}$, satisfying the $\Delta j = \Delta l = 3$ and
$\Delta K=2$ condition, are very close to each other at $\beta_{20}$ = 0.3.
Therefore the non-axial octupole $Y_{32}$ develops and with $\beta_{32}$
increasing from zero, an energy gap appears at $Z=98$.
Similarly, the spherical neutron orbitals $\nu 2g_{9/2}$ and $\nu 1j_{15/2}$
are very close to each other~\cite{Zhao2012_arxiv1209.6567}.
The neutron levels $[734]9/2$ originating from $1j_{15/2}$ and $[622]5/2$
originating from $2g_{9/2}$ are also close to each other and they just
lie above and below the Fermi surface.
This leads to the development of a gap at $N=150$ with $\beta_{32}$ increasing.
The $Y_{32}$ correlation in $N=150$ isotones
is from both the proton and the neutron and for $^{248}$Cf the correlation
is the most pronounced~\cite{Zhao2012_arxiv1209.6567}.

\section{Summary}
\label{sec:summary}

In this contribution we present some applications of the multi-dimensional
constrained covariant density functional theories in which
all shape degrees of freedom $\beta_{\lambda\mu}$ deformations
with even $\mu$ are allowed.
The potential energy surfaces of actinide nuclei in the $(\beta_{20}, \beta_{22}, \beta_{30})$
deformation space are investigated.
It is found that besides the octupole deformation, the triaxiality also
plays an important role upon the second fission barriers.
For most of even-even actinide nuclei, the triaxiality lowers the outer barrier
by 0.5 $\sim$ 1 MeV, accounting for about 10 $\sim$ 20\% of the barrier height.
The non-axial reflection-asymmetric $\beta_{32}$ shape in some transfermium
nuclei with $N=150$, namely $^{246}$Cm, $^{248}$Cf, $^{250}$Fm, and $^{252}$No
are studied.
Due to the interaction between a pair of neutron orbitals,
$[734]9/2$ originating from $\nu j_{15/2}$ and
$[622]5/2$ originating from $\nu g_{9/2}$, and
that of a pair of proton orbitals,
$[521]3/2$ originating from $\pi f_{7/2}$ and
$[633]7/2$ originating from $\pi i_{13/2}$,
rather strong non-axial octupole $Y_{32}$ effects have been found for $^{248}$Cf and
$^{250}$Fm which are both well deformed with large axial-quadrupole deformations,
$\beta_{20} \approx 0.3$.
For $^{246}$Cm and $^{252}$No, a shallow minima develops along
the $\beta_{32}$ deformation degree of freedom.

\section*{Acknowledgement}

%\begin{acknowledgement}
This work has been supported by
the 973 Program of China (2013CB834400),
Natural Science Foundation of China (10975100, 10979066, 11121403,
11175252, 11120101005, and 11275248),
and
Chinese Academy of Sciences (KJCX2-EW-N01 and KJCX2-YW-N32).
The results described in this paper are obtained on the ScGrid of
Supercomputing Center, Computer Network Information Center of Chinese Academy
of Sciences.
%\end{acknowledgement}

%\bibliographystyle{epj}
%\bibliography{../../../../information/refs/JabRef/sgzhou}
%\bibliography{nuclear,Nuclear-Fission}

%\begin{thebibliography}{}
%Author, Journal \textbf{Volume}, (year) page numbers
%Author, \textit{Book title} (Publisher, place year) page numbers
% etc
%\end{thebibliography}

\end{document}